\documentstyle[multicol,epsfig,aps,prl,array]{revtex}
\begin{document}
\draft

\title{Fluctuation and relaxation properties of pulled fronts:\\ a possible
scenario for non-Kardar-Parisi-Zhang behavior }

\author{Goutam Tripathy and Wim van Saarloos}
\address{Instituut--Lorentz, Universiteit Leiden, Postbus 9506,
  2300 RA Leiden, The Netherlands}
\date{\today} \maketitle

\begin{abstract}    
  We argue that while fluctuating fronts propagating into an unstable state
  should be in the standard KPZ universality class when they are {\em
pushed}, they 
  should not when they are {\em pulled}: The universal $1/t$ velocity
  relaxation of deterministic pulled fronts makes it unlikely that the KPZ
  equation is the appropriate effective long-wavelength low-frequency
  theory in this regime.  Simulations in 2$D$ confirm the proposed
  scenario, and yield exponents $\beta \approx 0.29\pm 0.01$, $\zeta
  \approx 0.40\pm 0.02$ for fluctuating pulled fronts, instead of the KPZ
  values $\beta=1/3$, $\zeta = 1/2$. Our value of $\beta$ is consistent
  with an earlier result of Riordan {\em et al.}
\end{abstract} 

\begin{multicols}{2}
Over a decade ago, Kardar, Parisi and Zhang (KPZ) \cite{kpz}
introduced their celebrated stochastic equation
\begin{eqnarray} \label{kpz}
{{\partial h }\over{\partial t}} &  = & \nu \nabla^2 h + {{\lambda}\over{2}}
(\nabla h)^2 + \eta ~,\\
\langle\eta({\bf x} ,t)\eta({\bf x}',t')\rangle & = & D \delta^d ({\bf x}-{\bf x'})
\delta (t-t') ~,\label{kpznoise}
\end{eqnarray}
to describe the fluctuation properties of growing interfaces with
height $h$ under the influence of  the  noise term $\eta$.
A clear ``derivation'' of the KPZ equation is  difficult to give, just 
as much as the Landau-Ginzburg-Wilson Hamiltonian can not
straightforwardly   be
``derived'' from the Ising model. However, one expects the KPZ
equation to be the proper effective long wavelength low frequency
theory for interfacial growth phenomena whose deterministic
macroscopic  
evolution equation is of the form 
\begin{equation}\label{macroscopic}
{{\partial h }\over{\partial t}} = v(\nabla h) + \mbox{curvature
  corrections}.
\end{equation}
Here $v(\nabla h ) $ is the deterministic growth velocity of a planar
interface as a function of the orientation $\nabla h$. For, as long
as the curvature corrections of the form $\nabla^2 h$ are nonzero, the 
long wavelength expansion of (\ref{macroscopic}) immediately yields
the gradient term in (\ref{kpz}). The philosophy is then that in the
presence of noise, all the relevant terms in the KPZ equation
(\ref{kpz}) are generated, and that this is sufficient to yield the asymptotic
KPZ scaling.  In agreement with
this picture, many  interface growth models have been found
\cite{krug1,halpinhealy,stanley,krug2} to show
the universal asymptotic scaling properties predicted by (\ref{kpz}).

A dynamical interface equation of the form (\ref{macroscopic}) is
appropriate for interfaces whose long wavelength and slow time
dynamics is essentially {\em local} in space and time, i.e., {\em dependent on
the local and instantaneous values of the slope and curvature}.  The
applicability of the KPZ equation is therefore not limited to
situations with a microscopically sharp interface: Many pattern
forming systems of the {\it reaction-diffusion} type  exhibit fronts whose
intrinsic 
width $l$ is finite. For curvatures $\kappa$ small compared to $l^{-1}$,
$\kappa l \ll 1$,  an effective interface approximation or
moving boundary approximation  of the form (\ref{macroscopic})
can then be derived using standard techniques \cite{examples}.  These approximations apply whenever the 
internal stability modes of the fronts relax exponentially on a short
time scale, so that an adiabatic decoupling becomes exact in the limit
$\kappa l \! \rightarrow \! 0$. The best known
example of such a type of analysis is for the curvature driven
growth in the Cahn-Hilliard equation, but moving boundary
techniques have recently been applied successfully to many other such
problems \cite{examples}. In all these cases, the internal relaxation
modes within the fronts or transition zones are indeed exponentially
decaying on a short time scale.

From the above perspective, recent results for the
relaxation properties of planar fronts propagating into an unstable state,
suggest an interesting new scenario for non-KPZ behavior. Fronts
propagating into unstable states generally come in two classes, so-called
{\em pushed} fronts and {\em pulled} fronts \cite{evs1}. Pushed
fronts propagating into an unstable state are the immediate analogue of
fronts between two linearly stable states.  In the thin interface
limit, $\kappa  l \ll 1$, the dynamics of such fronts becomes
essentially local and instantaneous, and   given by an equation of 
the form (\ref{macroscopic}); according to the arguments given above,
fluctuating pushed fronts should thus obey KPZ scaling:  following
standard practice by saying that the  $d$+1D  KPZ equation (where the
+1 refers to the time dimension) describes the fluctuations of a
$d$-dimensional interface,  the
conclusion is that fluctuations of $d$-dimensional pushed fronts in
$(d$+$1)$ {\em bulk dimensions} are described by the $d$+1D KPZ
equation. 

Pulled fronts, however, behave very differently from pushed ones. A
pulled front 
propagating into a linearly unstable state is basically ``pulled
along'' by the linear  growth dynamics  of small
perturbations spreading into the linearly unstable state. The crucial
new insight for our discussion is the recent 
finding  \cite{evs1,evs2} that pulled fronts {\em
can not be described} by an effective interface equation like
(\ref{macroscopic}) that is local and instantaneous in space {\em and}
time, even if they are weakly curved on the spatial scale. This just reflects the fact that the
dynamically important region of pulled fronts is the {\em semi-infinite}
leading edge region {\em ahead of the front}, not the nonlinear front
region itself. Technically, the breakdown of an interfacial
description is seen from the divergence of the 
solvability type integrals that arise in the derivation of a moving
boundary approximation in dimensions $d$$\ge$$2$
\cite{evs1}. More intuitively, the result can  be understood as follows: a
deterministic pulled front in $d$=$1$ relaxes to its asymptotic speed
$v^*$ with a universal power law as  $v(t)=v^*+c_1/t + c_{3/2}/t^{3/2}
+ \cdots$, where 
$c_1(<0)$ and $c_{3/2}$ are known coefficients
\cite{evs2}. Clearly, this very slow power law relaxation implies that
an 
adiabatic decoupling of the internal front dynamics and the large scale
pattern dynamics {\em can not be made} and hence that there is no long-wavelength
effective interface equation of the form (\ref{macroscopic}) for pulled
fronts.  There is then a
priori no reason to expect that fluctuating pulled fronts are in the
KPZ universality class!

It is our aim to test this scenario by
introducing a simple stochastic lattice model whose  front
dynamics can be changed from pushed to pulled by tuning a single
parameter. Our results are consistent with our conjecture that pulled
fluctuating fronts are not in the standard KPZ universality class, while pushed
fronts are. In fact, our results put an earlier empirical  finding of
Riordan {\em et al.} \cite{riordan} into a new perspective: These
authors obtained essentially the same growth exponent as we do for the
non-KPZ case, but the connection with the transition from
pushed to pulled front dynamics was not made.
\begin{figure}[tb]
\begin{center}
\leavevmode
\psfig{figure=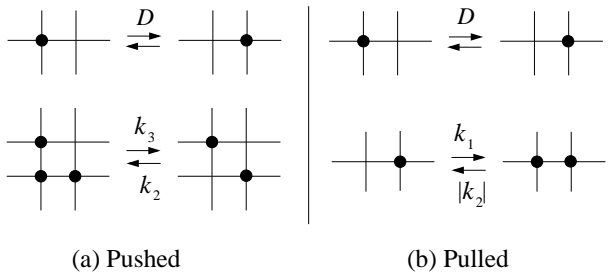,width=8cm}
\end{center}
\narrowtext
\caption{The two versions of our stochastic model for noisy pushed and
pulled fronts. The model is isotropic, i.e., all neighbors are probed
with equal probability. {\em (a)} The pushed case: Stochastic moves consist of
diffusive jumps of a particle to a neighboring empty site and birth and
death processes at sites whose two randomly chosen nearest neighbors are
occupied.  {\em (b)} The pulled case: The only difference with the pushed
case is in the birth and death processes. }
\label{fig:1.eps}
\end{figure}
Our stochastic model is motivated by  \cite{riordan}  and the results
for deterministic planar fronts in the nonlinear diffusion  equation
\begin{equation}
\partial \rho / \partial t = D \nabla^2 \rho + k_1 \rho + k_2 \rho^2
-k_3 \rho^3 ~.
\label{fkpp}
\end{equation}
As discussed in \cite{bj,evs1}, the planar fronts with $\rho>0$
propagating into the unstable state $\rho =0$ are pulled for all values
$k_2<\sqrt{k_1k_3/2}$ and pushed for larger $k_2$. In the pulled
regime, the asymptotic front velocity,  is $v^*=2\sqrt{Dk_1}$, while in
the pushed regime the asymptotic front velocity  equals $v^\dagger =
2\sqrt{Dk_1}[(-K+\sqrt{K^2+4})/32] $ where $K=k_2/\sqrt{k_1k_3}$. We
confine ourselves here to  
studying two limits where the stochastic front dynamics can easily be
understood 
intuitively.

We study the dynamics of particles on a square lattice, subject to the
constraints that no more than one particle can occupy each lattice
site. The stochastic moves are illustrated in Fig.\ 1. They consist of
diffusive hops of particles to neighboring empty sites and of birth and
death processes on sites neighboring an occupied site. In a mean field
approximation, this stochastic model is equivalent to a discrete version of
(\ref{fkpp}). We will study here the two cases indicated in Fig.~1. For
$k_1=0$ (Fig.~1a), planar fronts are definitely pushed: Since the linear
spreading speed $v^*=0$ for $k_1=0$, the front must then be pushed, even if
corrections to the mean field behavior are important in the front region or
behind the front. Likewise, when $k_3=0$ and $k_2<0$ (Fig.~1b), the
nonlinearities behind the front only limit the birth (growth) rate, so in
this limit the stochastic planar front is definitely pulled.
\begin{figure}[tb]
\begin{center}
\leavevmode
\psfig{figure=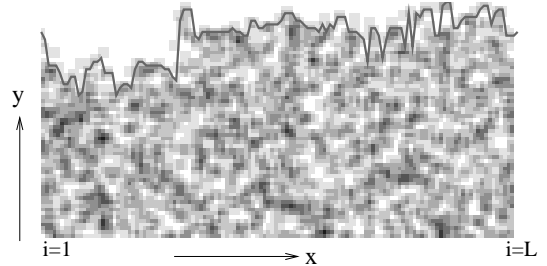,width=7cm}
\end{center}
\narrowtext
\caption[]{A snapshot of the coarse-grained density field ($m=1$).  The
interface position (continuous line) is obtained as the point where the
density crosses half its bulk value.}
\label{fig:fig2}
\end{figure}
Our simulations are done on 2D strips which are long in the $y$ direction
and of width $L$ in the $x$ direction. In the $x$ direction, periodic
boundary conditions are used. The Monte Carlo simulations are started with
a configuration in which the first few rows ($\approx 100$) of the lattice
are occupied with a probability equal to the equilibrium density. All other
lattice sites are empty. After an initial transient, the scaling properties
of the interface width are studied in the standard way using the following
definition of the interface height $h$. We define a coarse-grained density
variable at each lattice site as the average occupation of sites on a
$(2m+1)\times(2m+1)$ grid centered at that site. We then define the
position $h(x_i)$ of the interface as the first point where this
coarse-grained density reaches half the equilibrium density value.  Our
results for ensemble averaged width of the interface (see below) are
obtained by averaging over $100$ runs for the largest system $L=2048$ to
about $3200$ runs for the smallest $L=64$.  Although we have performed
simulations with $m=1,2$ and $m=3$ almost all the data presented subsequently are
those for a representative value of $m=2$. The coarse-grained density field
and the corresponding interface position $h$ for a typical configuration is
shown in Fig.~2.
\begin{figure}[tb]
\begin{center}
\leavevmode
\psfig{figure=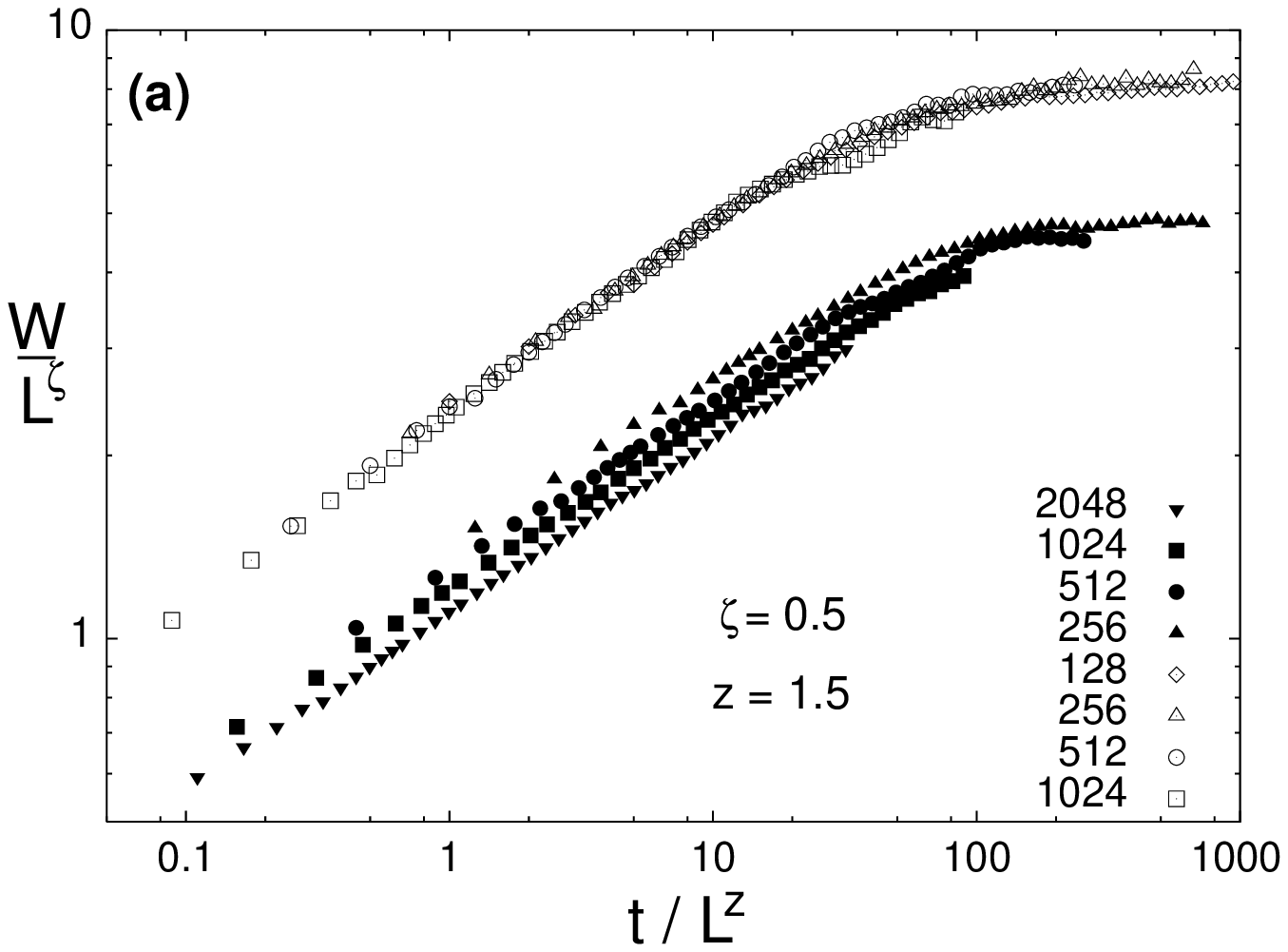,width=8cm}
\psfig{figure=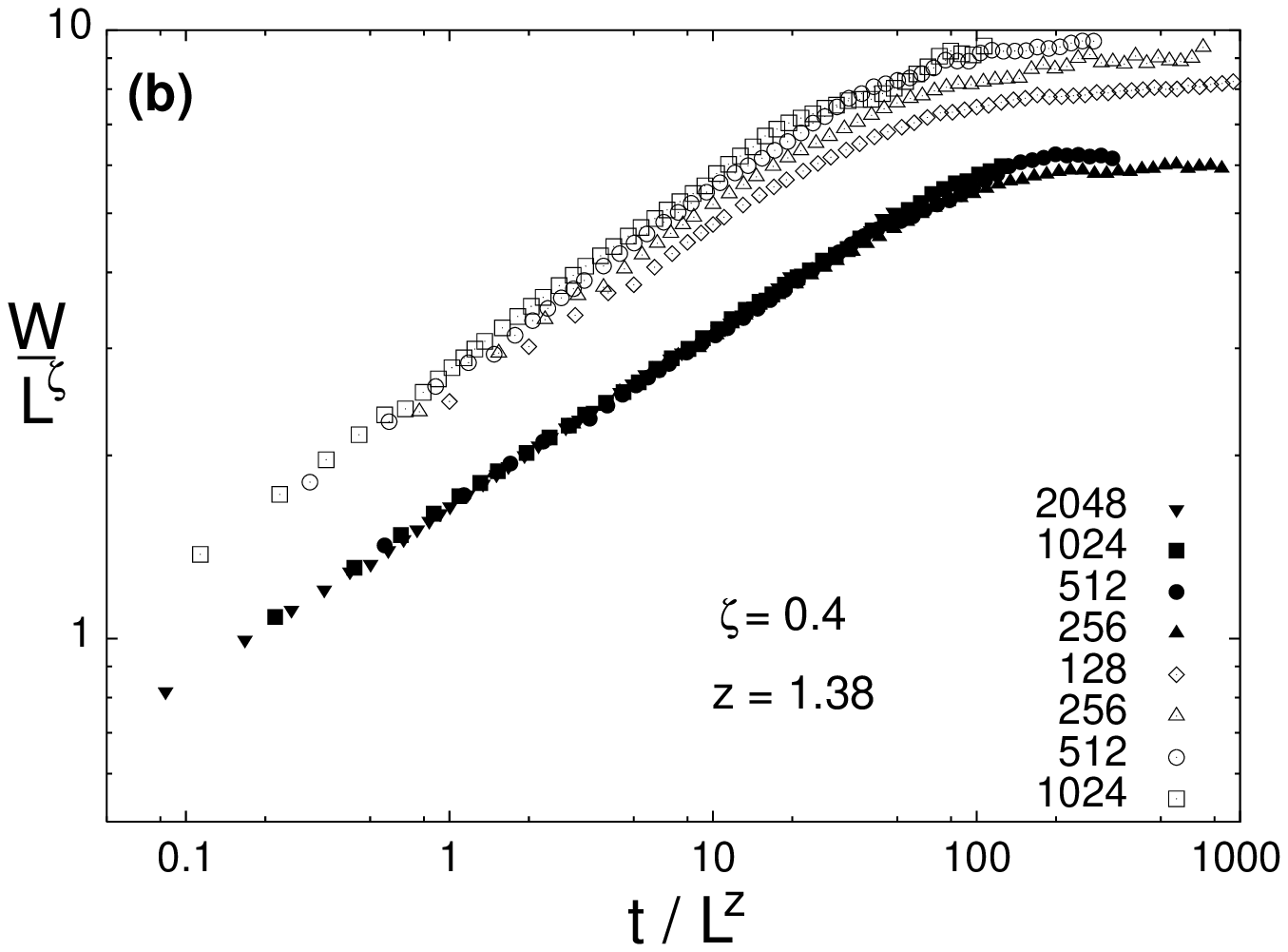,width=8cm}
\end{center}
\narrowtext
\caption{Scaling collapse of the width in the pushed (empty symbols) and
the pulled (solid symbols) cases using (a)
KPZ exponents $\zeta=0.5, z=1.5$ and (b) $\zeta=0.4, z=1.38$. The pushed
data values are multiplied by a factor $3$ for clarity.} 
\label{fig:push}
\end{figure}
The interface width $w$ of a given realization is defined in the usual way,
$w^2(t)=\overline{(h(x_i,t)-\overline{h(x_i,t)})^2}$, where the overbar
denotes a spatial average, $\overline{h}= L^{-1}\sum_{x_i} h(x_i,t)$. The
proper scaling to study is the ensemble averaged mean square interface
width $W^2 \equiv \langle w^2 \rangle$. As is well known, in the KPZ
equation $W$ obeys a scaling form $W(t)=t^{\beta} {\cal Y} \left (
{{t}\over{L^z}}\right)$. Here the scaling function ${\cal Y}(u)$ is about
constant for $u \ll 1$ and ${\cal Y} \sim u^{-\beta}$ for $u \gg 1$, with
the KPZ exponents $z=3/2 $ and $\beta =1/3$ in $1+1$D. For $t\gg L^z$, the
width saturates at $W_{sat} \sim L^\zeta$ where $\zeta=\beta z$ is the
roughness exponent. In Fig.\ 3, we show our data for stochastic pushed and
pulled fronts by plotting $W/L^{\zeta}$ versus $t/L^z$ for a range of
system sizes ($L=128$ to $2048$). Following standard practice, we always
plot the subtracted width $W^2(t)-W^2(0)$ to minimize the effect of
the initial
front width. The kinetic parameters are chosen to be $k_2=0.5, k_3=1.0$ for
the pushed model and $k_2=0, k_1=0.1$ for the pulled model. The diffusion
rate $D=0.25$ is the same in both cases. In Fig. 3a we use the 1+1D KPZ
exponents to obtain a data collapse. Clearly, good scaling collapse of the
pushed data confirms that the pushed fronts are in the universality class
of the 1+1D KPZ equation. By contrast, use of 1+1D KPZ values does not lead
to good scaling collapse of the pulled data.  In Fig. 3b we show the same
sets of data but now with exponents $\zeta=0.4$ and $z=1.38$ to obtain the
best possible scaling collapse of the pulled data. It is clear that the two
sets of exponents, though only moderately different from each other, are
well beyond error-bars. More accurate estimates of the exponents for the
pulled case were obtained as follows.  In Fig. 4 we fit a power law to the
non-saturated part of the width for the largest system $L=2048$ and obtain
$\beta\simeq 0.29\pm 0.01$ for the growth exponent. Plotting the saturated
width $W_{sat}$ as a function of system size $L$ (Fig. 4, inset) yields
$\zeta\simeq 0.4\pm 0.02$ for the roughness exponent. Once $\zeta$ is known
the dynamic exponent is obtained by requiring good scaling collapse of
Fig. 3b., $z=1.38\pm 0.06$.  The value of $\beta$ is consistent with that
reported by Riordan {\em et al.}  \cite{riordan} for this model, $\beta
=0.272 \pm 0.007 $, but their apparent value of $z\simeq 1$ is not the true
dynamic exponent related to the interface roughness through $\zeta=\beta
z$, since they studied the ensemble averaged width of the front
\cite{note}.
\begin{figure}[tb]
\begin{center}
\leavevmode
\psfig{figure=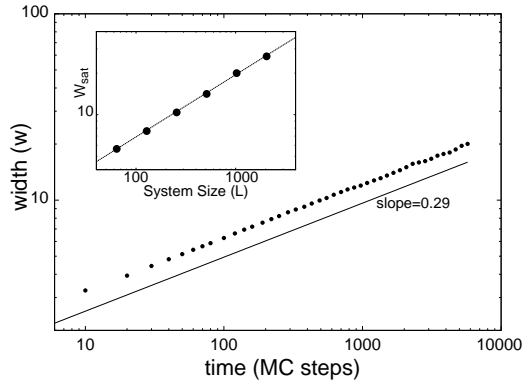,width=5cm,angle=-90}
\end{center}
\narrowtext
\caption{Scaling of the width in the pulled case for a system of
size $L=2048$. We have taken $k_1=0.1, k_2=-1.0$ and $D=0.25$.
The inset shows the saturated width
$W_{sat}$ vs $L$ plot on a log-log scale for 
$L=64\times 2^n (n=0,\cdots,5)$ and is consistent with $\zeta=0.4$
(solid line).} 
\label{fig:pull}
\end{figure}

Another way to investigate the possible difference with the 1+1D KPZ behavior
is to study the distribution $P(w^2/W^2)$. For
1D interface  models whose long time interface
configurations are given by a 
Gaussian distribution, like the KPZ model, the distribution function
$P(w^2/W^2)$ is uniquely determined, without adjustable parameters
\cite{racz}. As Fig.\ 5 shows, in the pushed regime our data are
completely consistent with this distribution function, but in the
pulled regime the measured distribution function deviates
significantly from the universal prediction for Gaussian interface
fluctuations. 
\begin{figure}[tb]
\begin{center}
\leavevmode
\psfig{figure=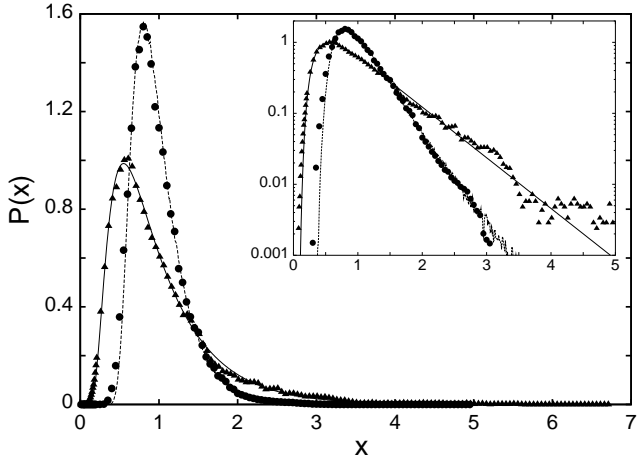,width=6cm,angle=-90}
\end{center}
\narrowtext
\caption[]{The probability distribution of the width of the interface
($x\equiv w^2/W^2$).  The triangles and the circles are for the pushed and
the pulled data of Fig.~3, respectively, and the solid and the dashed lines are the
universal distribution functions for the $1+1$D KPZ and $2+1$D KPZ equation,
 as obtained by R\'acz {\em et al.} \cite{racz}. Inset: the
same plot on a semi-log scale showing the agreement for large arguments.  }
\label{fig:wdist}
\end{figure}
The essential difference between pushed and pulled fronts is that for
pushed fronts the dynamically important region is the {\em finite}
transition zone between the two phases it separates, whereas for pulled
fronts it is the semi-infinite leading edge {\em ahead of the front itself}
\cite{evs1,evs2}. It is precisely for this reason that the wandering of
stochastic pulled fronts in one bulk dimension with multiplicative noise
was recently found to be subdiffusive and determined by the 1+1D KPZ
equation, not by a ``0+1D'' stochastic Langevin equation \cite{rocco}.  By
extending this idea it has been recently conjectured \cite{tripathy} that
the scaling exponents of stochastic pulled fronts in $d$+$1$ bulk
dimensions are generally given by the $(d$$+$$1)$$+$$1$D KPZ equation
instead of the $d$+1D KPZ equation, essentially because the dimension
perpendicular to the front can not be integrated out \cite{tripathy}. The
scaling exponents we find here in 2 bulk dimensions are indeed close to
those reported for the $2$+$1$D KPZ equation \cite{stanley}, the supposedly
most accurate values being $\zeta = 0.393(3)$, $\beta= 0.245(3)$
\cite{marinari}. Moreover, the probability distribution $P(w^2/W^2)$ of
pulled fronts fits the $P(w^2/W^2)$ of the 2+1D KPZ equation quite well
without adjustable parameters, see Fig. \ref{fig:wdist}.  For justification
and further exploration of this conjecture, we refer to \cite{tripathy}.

An interesting limit of our model is obtained when we further take
$k_2=0$ in Fig.~1b. In this case only birth and diffusion occurs, leading to an
equilibrium density $\rho=1$ behind the front. If we put $D=0$ as
well, the result is an Eden-like model 
\cite{stanley,krug2} with the modification that the probability
of adding a particle is proportional to the number of neighbors, not
independent of it. 
Numerical simulations in this Eden-like limit indicate that the KPZ exponents are
recovered, as it should, and hence that the model has a KPZ to non-KPZ
transition at intermediate values of $\rho_{eq}$ and $D$.

In conclusion, even though one should always be aware of the possibility of
a very slow crossover to asymptotic behavior in such studies \cite{brunet} --- a problem
that has plagued some earlier tests of KPZ scaling in, e.g., the Eden model
--- taken together our data as well as those of \cite{riordan} give, in our
opinion, reasonably convincing evidence for our scenario that the absence
of an effective interface description for deterministic pulled
fronts also entails non-KPZ scaling of 
stochastic pulled fronts.
 
We thank J. Krug, T. Bohr and Z. R\'acz for stimulating discussions. 
G. Tripathy is supported by the Dutch Foundation for  
Fundamental Research on  Matter (FOM).

\end{multicols}

\begin{references}
\bibitem{note} In \cite{riordan}, the width is not defined relative to
  $\overline{h}$; as a result, the long time data of \cite{riordan}
  show the diffusive wandering of $\overline{h}$, rather than the
  saturation of $W$.
\end{references}
\end{document}